\documentclass[a4paper,fleqn,usenatbib]{mnras}
\usepackage{newtxtext,newtxmath}

\usepackage[T1]{fontenc}
\usepackage{ae,aecompl}
\usepackage{comment}

\usepackage{graphicx}	
\usepackage{amsmath}	
\usepackage{epstopdf}
\usepackage{lineno}
\usepackage{xspace}
\usepackage{lineno}
\usepackage{multirow} 
\usepackage{longtable}
\usepackage{supertabular,booktabs}

\newcommand{\gama}{$\gamma$}
\newcommand{\fermi}{\textsl{Fermi}}

\title[\gama-ray variability in blazars]{Gamma-ray Blazar variability: New statistical methods of time-flux distributions}

\author[Duda\& Bhatta]{
Jaros{\l}aw Duda$^{1}$\thanks{E-mail: dudajar@gmail.com} \ \&
Gopal Bhatta$^{2}$\thanks{E-mail: gopal.bhatta@ifj.edu.pl}
\\
$^{1}$Institute of Computer Science and Computer Mathematics, Jagiellonian University, ul. Lojasiewicza 6,
 30-348 Krak\'ow, Poland\\
$^{2}$Institute of Nuclear Physics Polish Academy of Sciences, PL-31342 Krak\'ow, Poland
}

\date{Accepted XXX. Received YYY; in original form ZZZ}

\pubyear{2015}

\begin{document}
\label{firstpage}
\maketitle

\begin{abstract}

Variable \gama-ray emission from blazars, one of the most powerful classes of astronomical sources featuring relativistic jets, is a widely discussed topic. In this work, we present the results of a variability study of a sample of 20 blazars using \gama-ray (0.1--300~GeV) observations from Fermi/LAT telescope. Using maximum likelihood estimation (MLE) methods, we find that the probability density functions that best describe the \gama-ray  blazar flux distributions use the stable distribution family,  which generalizes the Gaussian distribution.  The results suggest that the average behavior of the \gama-ray flux variability over this period can be characterized by log-stable distributions. For most of the sample sources, this estimate leads to standard log-normal distribution ($\alpha=2$). However, a few sources clearly display heavy tail distributions (MLE leads to $\alpha<2$), suggesting underlying multiplicative processes of infinite variance. Furthermore, the light curves were analyzed by employing novel non-stationarity and autocorrelation analyses.  The former analysis allowed us to quantitatively evaluate non-stationarity in each source --- finding the forgetting rate (corresponding to decay time) maximizing the log-likelihood for the modeled evolution of the probability density functions. Additionally, evaluation of local variability allows us to detect local anomalies, suggesting a transient nature of some of the statistical properties of the light curves.  With the autocorrelation analysis, we examined the lag dependence of the statistical behavior of all the $\{(y_t,y_{t+l})\}$ points, described by various mixed moments, allowing us to quantitatively evaluate multiple characteristic time scales and implying possible hidden periodic processes.
\end{abstract}



\section{Introduction }
\label{sec:intro}
The class of active galaxies that emit profusely at radio frequencies are
radio-loud galaxies. These galaxies often show the
presence of kiloparsec (kpc) scale relativistic jets. If the jet is oriented
towards the Earth, the relativistic effects become dominant such that
the Doppler-boosted non-thermal emission makes the sources remarkably
brighter over a wide range of electromagnetic frequencies. The
emission is found to be more pronounced at higher energies,
e.g., X-ray and \gama-rays. These objects could also be the sources of
neutrinos flying through the inter-galactic
medium\citep[see][]{2018Sci...361.1378I,2018Sci...361..147I}.  In
addition, the kpc scale jets seem to be most efficient cosmic
particle accelerators, wherein the particles, mainly leptons, are
accelerated to several orders of rest-mass electron energies. 
As a result, large amounts of accelerated high-energy particles become 
sources of incoherent synchrotron emission by decelerating into the
ambient jet magnetic field, thereby making the extended jet
``visible". These energetic particles might also up-scatter the
surrounding synchrotron photons which they themselves produced
\citep[see][]{Maraschi1992,Mastichiadis2002} or low-energy
electrons of external origins e.g., from the accretion disk
(\citealt{Dermer1993}), broad-line region (\citealt{Sikora1994}), and
dusty torus (\citealt{Blazejowski2000}), resulting in a large output
of MeV--TeV emission.

 Variability over minute to decade timescales is one of the
 characteristic, defining properties of blazars.  Numerous studies in
 various energy bands and across all timescales have been conducted over the
 years using all available ground and space based instruments
 \citep[see][]{2019Galax...7...28R,Madejski2016,Bhatta2018}.
 Particularly in the \gama-ray regime, studies of power density spectra have
 shown that the statistical nature of the variability can well be
 described by a single power-law in the Fourier domain \citep[see][and references therein]{Bhatta2020}; in some sources,
 applying continuous autoregressive models leads to the inference of breaks in the power
 spectra, possibly corresponding to characteristic timescales
 \citep{2019ApJ...885...12R}. Indeed, time domain
 analysis of blazars serves as one of the most important tools to
 unravel the physical process occurring in the innermost regions
 around the central engines.  The aim of this current work is to
 explore the statistical properties of the light curves in order to
 infer more fundamental mathematical properties of the process(es) driving
 the variability, such as linearity and stationarity. Moreover, as
 blazar variability timescales span a wide temporal range, i.e., from
 a few minutes to several decades, it is natural to conceive of
 the observed (total) flux variability as a combination of flux
 variability owing to individual stochastic processes occurring within the
 different sub-volumes of the parsec-scale accretion disk and
 kpc-scale relativistic jets. In such a scenario, it is an
 important question to ask whether such a combination is of an additive
 or a multiplicative nature. Several recent works on blazars show that
 the blazar flux distribution is well representated by a heavy
 tailed log-normal PDF.  Particularly, the \gama-ray fluxes of some of the
 brightest blazars have been found to follow log-normal distributions
 \citep[see e. g.][and references
   therein]{Bhatta2020,Shah2018}. Such flux log-normality is often
 interpreted as an indication of the non-linearity in the
 multiplicative processes.  In the context of AGN, it has been
 proposed that long-memory processes, such as flicker noise,
 originate due 
to the inward propagation of fluctuations in the
 mass accretion rates, which in turn create the rapid variability near
 the central region \citep[see][]{Lyubarskii1997}. 
Magnetohydrodynamic (MHD) simulations of a thin disk
 around the black hole results in the observed log-normality along with
 the linear RMS-flux relations \citep{Hogg2016}.  Similarly, in the
 statistical model of minijets-in-a-jet {\citep[see][]{2009MNRAS.395L..29G} isotropically oriented
 Doppler-boosted mini-jets are distributed over the span of the bulk
 relativistic flow.  The total flux from the whole emission region is
 then of the log-normal form \citep{Biteau 2012}.

The skewness of the flux distribution suggests that the variability
stems from multiplicative processes, which are associated in some
models with the accretion disk. In this paper, we study the rms-flux relation and
emphasize its link to Pareto distributions. The minijets-in-a-jet
statistical model reconciles the fast variations and the statistical
properties of the flux of blazars at very high energies.

As an attempt to understand the phenomenon of multi-timescale,
multi-frequency variability in the sources, several emission models
have been invoked; some of the widely discussed models include various
magnetohydrodynamic instabilities in the turbulent jets
\citep[e.g.][]{bhatta13, Marscher14}, shocks traveling down jets
\citep[e.g.][]{Marscher1985,Spada2001}, the aforementioned jets-in-a
jet model \citep{2009MNRAS.395L..29G} and effects of jet orientation
or geometric models \citep[e.g.][]{2016MNRAS.461.3047L}. In spite of
the collaborative efforts across many instruments and observations,
and in modeling and theory, the details of the processes shaping
multi-timescale variability still remains debated.  The importance of
time domain analysis with a focus on constraining the nature of the
variability can not be exaggerated as variability studies provide us
with an excellent tool to probe the energetics of supermassive black
hole systems.

In this work, we perform a statistical analysis of decade-long \textit{Fermi}-LAT
observations of 20 blazars that were presented in
\citet{Bhatta2020}. The source names, their 3FGL catalog names, source
classifications, r.a., declinations, and redshifts are presented in columns 1, 2, 3,
4, 5 and 6, respectively, of Table \ref{table:1}. In Section
\ref{sec:2}, the details of the analyses methods carried out on the
\gama-ray light curves are discussed.  The results and the discussions
are presented in Section \ref{sec:3}.

\begin{table*}
        \caption{General information about the sample \textit{Fermi}LAT blazars}
        \centering
        \label{table:1}
        \begin{tabular}{l|l|l|l|l|c}
                \hline
                Source name & Source class &3FGL name&R.A. (J2000) & Dec. (J2000) & Redshift ($z$) \\
                \hline
    W Comae 	&	3FGL J1221.4+2814	&	BL Lac 	&	 $12^{\rm h}~21^{\rm m}~31\fs7$ 	&	 $+28^\circ~13\arcmin~59\arcsec$   &	0.102	 \\
    PKS 1502+106&	3FGL J1504.4+1029	&	FSRQ 	&	 $15^{\rm h}~04^{\rm m}~25\fs0$ 	&	 $+10^\circ~29\arcmin~39\arcsec$   &	1.84	 \\
    4C+38.41 	&	3FGL J1635.2+3809 	&	FSRQ 	&	 $16^{\rm h}~35^{\rm m}~15\fs5$ 	&	 $+38^\circ~08\arcmin~04\arcsec$   &	1.813	 \\
    BL Lac 	&	3FGL J2202.7+4217 	&	BL Lac 	&	 $22^{\rm h}~02^{\rm m}~43\fs3$ 	&	 $+42^\circ~16\arcmin~40\arcsec$   &	0.068	 \\
    3C 279 	&	3FGL J1256.1$-$0547 	&	FSRQ 	&	 $12^{\rm h}~56^{\rm m}~11\fs1665$ 	&	 $-05^\circ~47\arcmin~21\farcs523$ &	0.536	 \\
    CTA 102 	&	3FGL J2232.5+1143 	&	FSRQ 	&	 $22^h32^m36.4^s$ 	&	 $+11^\circ~43\arcmin~51\arcsec$ 	&	1.037	 \\
    4C +21.35 	&	3FGL J1224.9+2122 	&	FSRQ 	&	 $12^h24^m54.4^s$ 	&	 $+21^\circ~22\arcmin~46\arcsec$ 	&	0.432	 \\
    Mrk 501 	&	3FGL J1653.9+3945 	&	BL Lac 	&	 $16^h53^m52.2167^s$ 	&	 $+39^\circ~45\arcmin~36\farcs609$ 	&	0.0334	 \\
    PKS 0454$-$234&	3FGL J0457.0$-$2324	&	BL Lac 	&	 $04^h 57^m03.2^s$ 	&	 $-23^\circ~24\arcmin~ 52\arcsec$ 	&	1.003	 \\
    1ES 1959+650 &	3FGL J2000.0+6509 	&	BL Lac 	&	 $19^h59^m59.8521^s$ 	&	 $+65^\circ~08\arcmin~54\farcs652$ 	&	0.048	 \\ 
    PKS 1424$-$418&	3FGL J1427.9$-$4206 	&	 FSRQ	&	 $14^h27^m56.3^s$ 	&	 $-42^\circ~06\arcmin~19\arcsec$ 	&	1.522	 \\
    PKS 2155$-$304&	3FGL J2158.8$-$3013 	&	BL Lac 	&	 $21^h58^m52.0651^s$ 	&	 $-30^\circ~13\arcmin~32\farcs118$ 	&	0.116	 \\
    S5 0716+714 &	3FGL J0721.9+7120 	&	BL Lac 	&	 $07^h21^m53.4^s$ 	&	 $+71^\circ~20\arcmin~36\arcsec$ 	&	0.3	 \\
    3C 66A 	&	3FGL J0222.6+4301 	&	BL Lac 	&	 $02^h22^m41.6^s$ 	&	 $+43^\circ~02\arcmin~35\farcs5$ 	&	0.444	 \\
    Mrk 421 	&	3FGL J1104.4+3812 	&	BL Lac 	&	 $11^h04^m273^s$ 	&	 $+38^\circ~12\arcmin~32\arcsec$ 	&	0.03	 \\
    ON +325 	&	3FGL J1217.8+3007 	&	BL Lac 	&	 $12^h17^m52.1^s$ 	&	 $+30^\circ~07\arcmin~01\arcsec$ 	&	0.131	 \\
    AO 0235+164 &	3FGL J0238.6+1636	&	BL Lac 	&	 $02^h 38^m38.9^s$ 	&	 $+16^\circ~ 36\arcmin~59\arcsec$ 	&	0.94	 \\
    PKS 1156+295&	3FGL J1159.5+2914 	&	BL Lac 	&	 $11^h59^m31.8^s$ 	&	 $+29^\circ~14\arcmin~44\arcsec$ 	&	0.7247	 \\
    3C 454.3  	&	3FGL J2254.0+1608 	&	FSRQ 	&	 $22^h53^m57.7^s$ 	&	 $+16^\circ~08\arcmin~54\arcsec$ 	&	0.859	 \\
    3C 273 	&	3FGL J1229.1+0202 	&	FSRQ 	&	 $12^h29^m06.6997^s$ 	&	 $+02^\circ~03\arcmin~08\farcs598$ 	&	0.158	 \\
                \hline
        \end{tabular}
\end{table*}

\section{Observations and data processing}
\label{obs}

The \gama-ray observations of the sample blazar were obtained from the
Large Area Telescope (LAT) onboard the \textit{Fermi Gamma-ray Space Telescope}
(\fermi) \citep{Atwood2009}.  The telescope has a large effective area
($> 8000\ cm^{2}$) to collect high energy photons coming from a wide
field of view ($>$ 2 sr). Moreover, the instrument can resolve
astronomical sources with a high angular resolution, that is, ($ <
3.5^{\circ}$ around 100~MeV and $< 0.15^{\circ}$ above 10~GeV).  To construct
the source light curves, Pass 8 data fom the \fermi/LAT 3FG catalog
were processed using Fermi Science
Tools{\footnote{\url{https://fermi.gsfc.nasa.gov/ssc/data/analysis/software/}}}
and following the standard procedures of the unbinned likelihood
analysis\footnote{\url{https://fermi.gsfc.nasa.gov/ssc/data/analysis/scitools/likelihood_tutorial.html}}. In
particular, the photon events in the energy range 0.1--300 GeV
classified as ``evclass=128, evtype=3'' were considered. A circular
region of interest (ROI) of $10^{\circ}$ radius centered around each
source was chosen; also, the zenith angle was limited to $ <$
90$^{\circ}$ in order to minimize the contamination from the Earth.
The Fermi Science Tools were used to perform analysis using the
\fermi/LAT 3FG catalog, Galactic diffuse emission model and isotropic
model for point sources. Moreover, the Galactic and extra-galactic
diffuse $\gamma$-ray emission models \emph{ gll\_ iem v06.fit} and
\emph{ iso\_P8R2 SOURCE V6 v06.txt} were also incorporated. To
generate the weekly-binned light curves, a maximum-likelihood analysis,
using the task \emph{ gtlike}, was performed on the photon events, and
test statistics $\geq$ 10 (equivalently $\gtrsim 3 \sigma$)
\citep{Mattox1996} were considered. For details on the data processing,
refer to \citet{Bhatta2020}.

\section{Methodology and analysis }
\label{sec:2}

Following the extended methodology as explained in  \cite{Duda2018},
we first normalized marginal distributions with a parametric
distribution (log-stable here) as in copula theory~\citep{copula}, and
then modeled the evolution of normalized variables, or joint distribution
for autocorrelations. In both the cases, the PDFs were represented in terms of a polynomial
basis.

\subsection{Additive and multiplicative processes}

In an additive process, the observed flux, say $X$, can be considered the
sum of the fluxes produced at a number of randomly distributed
smaller emission regions, i.e., $X=\sum_{i=1}^{N}x_{i}$. If the
number of processes, assumed to be independent and identically
distributed (i.i.d.) throughout the bulk emission region, becomes very large,
i.e., $N \rightarrow \infty$, then by the central limit theorem, the total
flux tends to follow a normal distribution. This distribution can also
originate in stationary and linear systems with finite
moments. In general, autoregressive processes, damped random walks,
shot noise, and Brownian motion can be described as linear and additive
processes. On the other hand, if the observed flux results from the
multiplication of a large number of smaller fluxes,
i.e., $X=\prod_{i=1}^{N} x_{i}$, then the integrated flux can follow
a highly-skewed heavy tailed log-normal distribution. Such multiplicative
processes are then ascribed to the non-linearity of the
system. Multiplicative processes, e.g., the Volterra process
\citep[see][]{Priestley1988}, are widely discussed in the literature,
such as in biological contexts by \citet{Mitzenmacher2004}, and in
financial time series analysis by \citet{Zanette2020}.

\subsection{Normalization with log-stable distribution}
\begin{figure*}
\begin{center}
\includegraphics{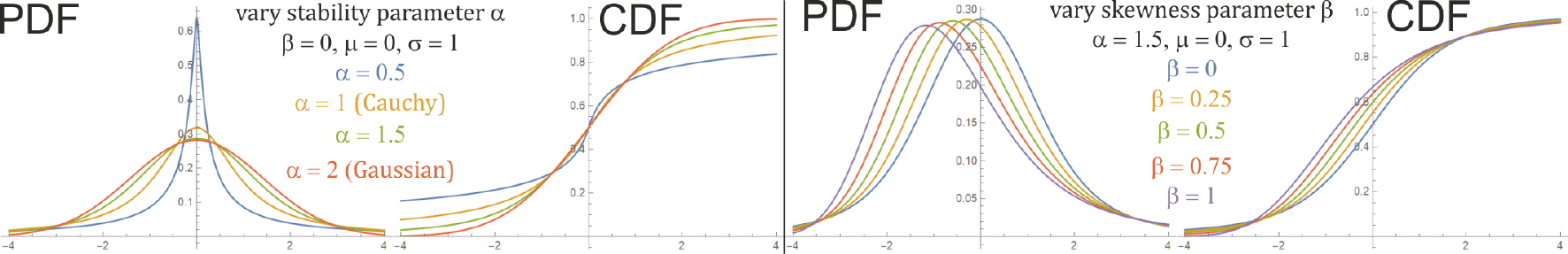}
\caption{Some of the model probability distribution functions (PDF) and cumulative
  distribution functions (CDF) for stable distributions that are employed here to explore the  $\gamma$-ray flux (logarithmized) distribution of the sample blazars.  For the maximal value $\alpha=2$, it represents the
  normal distribution (Gaussian), and the skewness parameter $\beta$
  has no effect.  For $\alpha<2$, such a distribution has heavy tails
  following $\sim |x|^{-\alpha-1}$, and therefore posseses an  infinite
  variance. According to the generalized central limit theorem \citep{Gnedenko},
  the sum of a number of random variables with symmetric
  ($\beta=0$) distributions having power-law tails decreasing as
  $|x|^{-\alpha-1}$, where $0 < \alpha < 2$, will tend to be a stable
  distribution.}
\label{fig:0}
\end{center}
\end{figure*}

\begin{figure*}
\begin{center}
\includegraphics[width=0.95\textwidth]{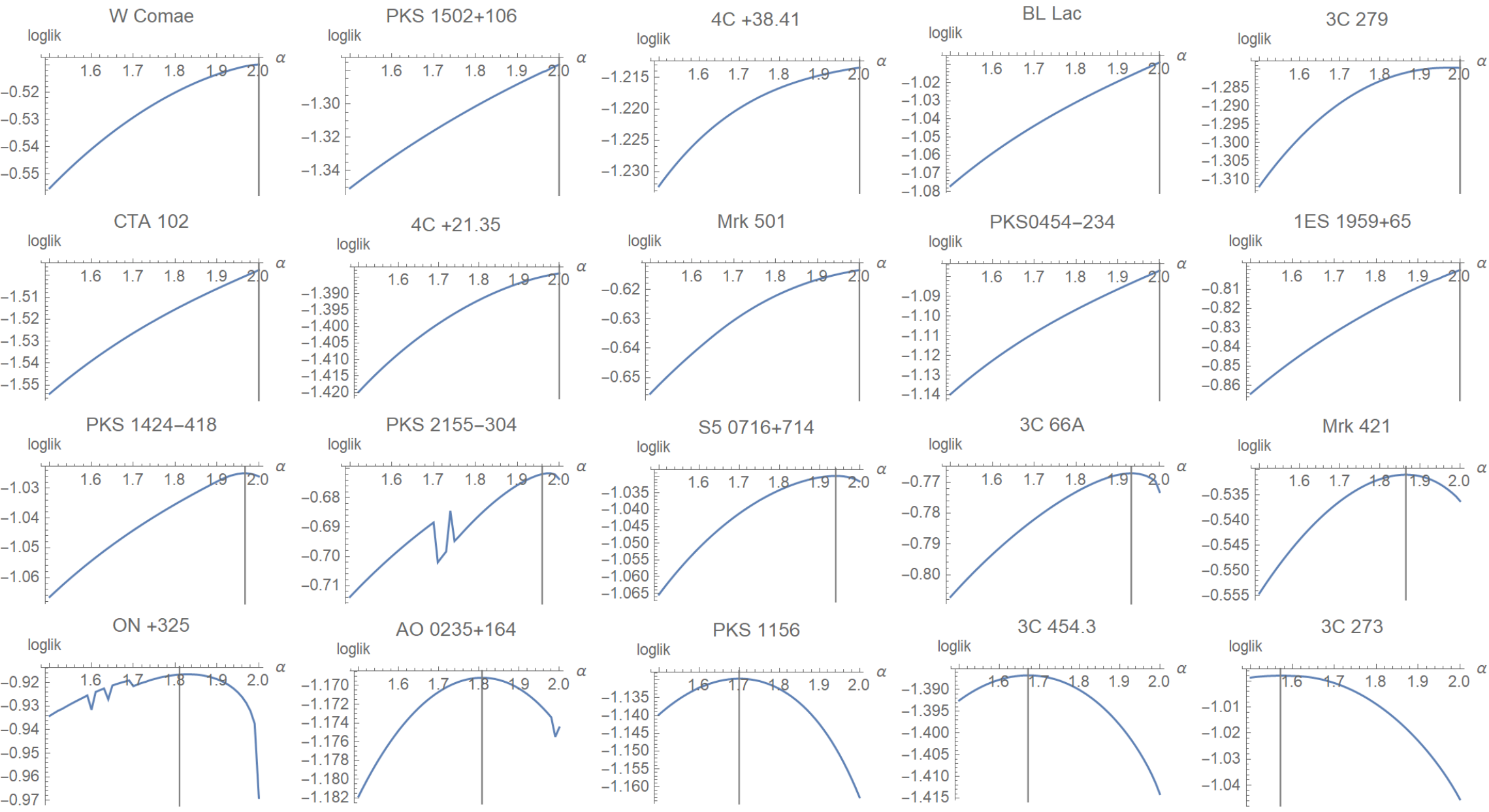}
\caption{MLE Log-likelihood evaluations for $(\ln(x_t))$ sequences using
  the stable distribution with various fixed values of the $\alpha$ parameter for
  all 20 objects. Discontinuities correspond to large changes in
  the optimal value of $\beta$.  The gray lines denote the fit yielding the most likely value of $\alpha$, which can be used to characterize a  given object e.g., for classification, and in particular quantifying the tail   type of its distribution.  We can see that in particular the last five objects
  clearly lead to $\alpha<2$, suggesting infinite-variance heavy  tails.   }
\label{fig:1}
\end{center}
\end{figure*}

As in copula theory, for the methodology used here, it is convenient
to first normalize flux values $(x_t)$ to $y_t=f(x_t)\in[0,1]$ having
a nearly uniform distriubtion in the $[0,1]$ range (furthermore,
Fig.~\ref{fig:4} presents these normalized values $(y_t)$).  Here, $f$
ideally is the cumulative distribution function (CDF) that this sample
comes from,  and should represent the probability density averaged over
the entire time period ($\approx 10$ years here). While this
normalization could be performed by just sorting the values and
assigning positions in order (the so-called empirical distribution
function), using a parametric family would give a better understanding
and suggest a universal behavior.  In non-stationarity analysis we
will additionally search for evolution of the probability density
during this time period, as a correction to density used for
normalization. In autocorrelation analysis, for pairs of values
shifted by various lags, we will evaluate distortions from the uniform
joint distribution on $[0,1]^2$.

A standard assumption for the parametric distribution of this type of
data, suggested by the central limit theorem for multiplicative
processes, is the log-normal distribution: a Gaussian distribution for
logarithmized values. To verify this assumption, we tested two larger
families containing Gaussian distributions: exponential power
distributions, $\rho(x)\sim \exp(-|x|^\kappa)$, and stable
distributions also containing heavy tails, $\sim |x|^{-\alpha-1}$
for $\alpha<2$. The highest log-likelihoods were achieved by
using stable distributions for logarithmized values; hence they were
applied for normalization (these evaluations are presented in
Fig.~\ref{fig:2}).

The stable distribution \citep{stable} is defined by four parameters: $\mu,
\sigma, \alpha, \beta$. As in the Gaussian distribution, it has 
a location parameter, $\mu\in (-\infty,\infty)$, and a scale parameter, 
$\sigma\in(0,\infty)$. Additionally it has a stability   parameter, $\alpha\in (0,2]$. For $\alpha=2$ we get the standard Gaussian distribution; for   $\alpha=1$ we get the Cauchy distribution with 
heavy tails following $1/x^2$. Generally for $\alpha\in (0,2)$ it has $|x|^{-\alpha-1}$
  heavy tails, leading to infinite variance. This family also has a
skewness parameter, $\beta\in[-1,1]$, 
which allows for some asymmetry
  in the distribution. However, its influence weakens when $\alpha$
  approaches 2, and for $\alpha=2$ this parameter has no
  effect. Examples of probability distribution functions (PDF) and
  cumulative distribution functions (CDF) for some combinations of parameters of the
  stable distribution are presented in
  Fig.~\ref{fig:0}. It can be defined using
    the characteristic function $\varphi$:}

$$\rho_{\mu\sigma\alpha\beta}(x)=\frac{1}{2\pi}\int_{-\infty}^\infty \varphi_{\mu\sigma\alpha\beta}(t)\, e^{-ixt}dt$$
\begin{equation} \varphi_{\mu\sigma\alpha\beta}(t)=\exp\left(i t \mu -|\sigma t|^{\alpha}(1-i\beta\, \textrm{sgn}(t)\tan(\pi \alpha/2)\right) \end{equation}

As the name suggests, these distributions have additional universality
which might suggest a hidden mechanism --- they are stable as per the
central limit theorem, this time in its generalized version
\citep{Gnedenko}. While addition of finite-variance i.i.d. random variables
asymptotically leads to the Gaussian distribution, for
infinite-variance variables such a summation usually leads to a stable
distribution (at least for $\beta=0$). For some of the sources in our
sample --- those with $\alpha=2$ --- there is good agreement with the
log-stable distribution, suggesting a multiplicative process with
finite variance; however the variance is infinite for the remaining
sources (those with $\alpha<2$).

\begin{figure*}
\begin{center}
\includegraphics[width=0.95\textwidth]{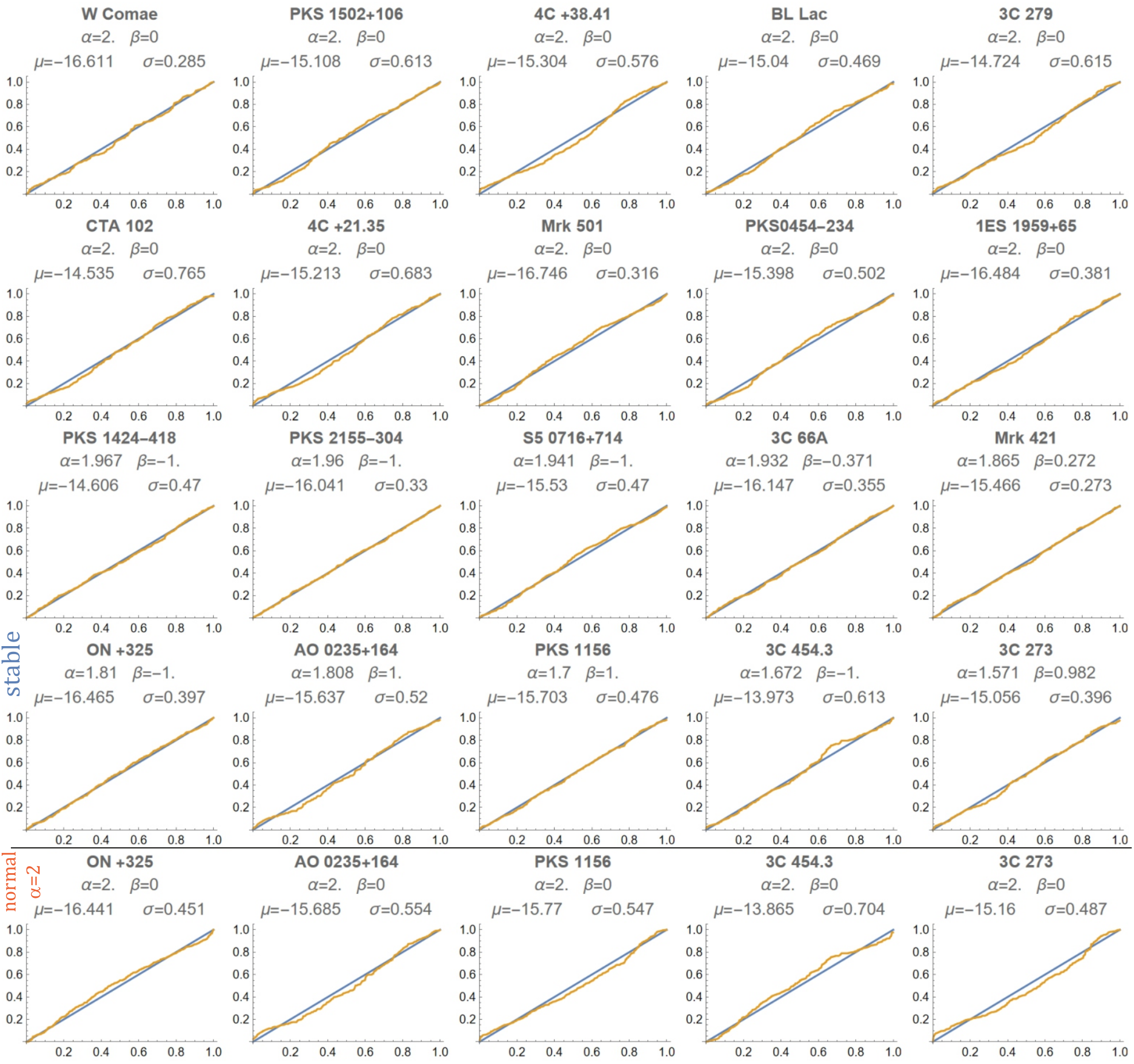}
\caption{Visual evaluation of the level of agreement of MLE stable
  distributions for the 20 observed time series' cumulative
  distribution functions (CDFs); the orange curve being equal to the
  blue diagonal would mean perfect agreement.
Specifically, the original time series
  $(x_t)$ was first logarithmized, then we performed an MLE
  (the most likely values of $\alpha,\beta, \mu, \sigma$ parameters are
  written in each panel). Then, 
  the sequence $y_t=\textrm{CDF}_{\alpha\beta\mu\sigma}(\ln(x_t))$
  was calculated using the CDF
  for parameters optimized for a given sequence. The orange curves
  are the sorted values of ${y_t}$, ideally from the uniform distribution which
  would yield the diagonal (blue). We can see that agreement between  the curves is quite
  decent; the $\alpha=2$ cases correspond to just the log-normal
  distribution. However, as seen in Fig.~\ref{fig:1}, some objects
  have clearly lower values of $\alpha$, denoting heavier tails. The $\beta$
  parameter denotes asymmetry and is limited to $[-1,1]$, which seems
  insufficient for a few sequences. The additional last row contains the
  same sources as the previous row, but normalized with log-normal
  distributions; we can see a larger discrepancy between the orange and blue
  curves. Disagreements near 0 and 1 suggest improper assumptions 
   on the tail
    behavior of the probability density; disagreements near the center
    center suggest an improperly assumed body of the probability
    distribution}.
\label{fig:2}
\end{center}
\end{figure*}

In the analysis presented here, the flux values $(x_t)$ were first
logarithmized, then for each individual object we performed a
maximum likelihood estimation (MLE) of parameters of the stable
distribution using Wolfram Mathematica software. To verify estimation
of $\alpha$ and evaluate its accuracy, we also performed
estimation with various fixed values of $\alpha$; those
log-likelihoods are presented in Fig.~\ref{fig:1}. We can see that for
some objects these fits suggest $\alpha<2$ and heavy tails, especially
in 3C~273, 3C~454.3, PKS~1156+295, AO~0235+164, and ON +325.

We then performed normalization using cumulative distribution
functions (CDF) of the most likely distributions: assuming a given sequence
$(\ln(x_t))$ has led, by MLE, to parameters $(\alpha,\beta,\mu,\sigma)$, 
we calculated the sequence
\begin{equation}
  y_t = \textrm{CDF}_{\alpha\beta\mu\sigma} (\ln(x_t))
\end{equation}
which would be from the uniform distribution on $[0,1]$ if $(\ln(x_t))$ was exactly from this stable distribution.

Beside log-likelihood tests, we also performed a visual
evaluation to test if such normalized variables $(y_t)$ are from a nearly
uniform distribution: by sorting them (empirical distribution) and
comparing with the diagonal, which would be obtained for the uniform
distribution. Fig.~\ref{fig:2} presents such a visual evaluation,
where we can see a relatively good agreement, especially at the
boundaries corresponding to tails.

In this Figure, we also list the stable distribution parameters
$\alpha, \beta, \mu, \sigma$ found via MLE. For most of the sequences
we obtained $\alpha=2$, which means that indeed the log-normal
distribution has turned out to be the best choice. However, the last
few sequences in this Figure (they were ordered by $\alpha$) yielded
lower values of $\alpha$ in this ML estimation, suggesting heavier
tails. Parameters of such a ML estimation can be treated as features
of objects, e.g., for classification purposes, especially the $\alpha$
parameter defining the type of tail of the distribution.

The normalized sequences $(y_t)$ are later presented in
Fig.~\ref{fig:4} as dots, where we can see that in the horizontal
direction they have a nearly uniform distribution. However, the local
density evolves in the vertical direction corresponding to time, as is
considered in non-stationarity analysis.

\subsection{Modelling non-stationarity with polynomials of evolving contribution}

After normalization, the variables $(y_t)_{t=1..n}$ are from 
nearly-uniform distributions. Here we would like to model any distortions from
this uniform distribution, such as its evolution for non-stationarity
analysis, by representing this density as a polynomial and modelling its
coefficients as discussed in \cite{Duda2018}.

For this purpose we could model the joint distribution of the pairs $\{(y_t,t)\}$,
with times $t$ rescaled to the range $[0,1]$, and predict the
conditional distributions $\rho(y|t)$ using a polynomial model for their joint distributions. We performed 10-fold cross-validation tests of log-likelihood for such an approach, but it led to an inferior
evaluation compared to an adaptive approach; hence we will focus only on
the adaptive approach here, especially since it also provides evaluation of
non-stationarity of the sequences.

We would like to model distortions from uniform density, $[0,1]$, 
(for normalized variables) as linear combinations using some basis $\{f_j:j\in B\}$, $B^+=B\backslash\{0\}$, $f_0=1$:
\begin{equation}
  \rho(y) = \sum_{j\in B} a_j\ f_j(y)= 1 +\sum_{j\in B^+} a_j\ f_j(y).
\end{equation}
As discussed in \cite{Duda2018}, these coefficients have similar
interpretations as moments: $a_1$ as the expectation value, $a_2$ as the
variance, $a_3$ as skewness, $a_4$ as kurtosis, etc. Using the orthornomal
family of functions $\int_0^1 f_i(y) f_j(y) dy=\delta_{ij}$,
the mean-square error estimation is given by just averages of functions
over the data sample $(y_i)_{i=1..n}$:
\begin{equation}\label{est}
  a_j = \frac{1}{n}\sum_{i=1}^n f_j(y_i)
\end{equation}

We tested various orthornormal families including the trigonometric family, and
generally the best results were obtained for (rescaled Legendre)
polynomials: $f_0,f_1,f_2,f_3,f_4$ are correspondingly:
$$ 1,\sqrt{3}(2y-1), \sqrt{5}(6y^2-6y+1),$$ $$\sqrt{7}(20y^3-30y^2+12y-1), \ \text{and} \ (70y^4-140y^3+90y^2-20y+1)$$

For adaptivity we can replace the average in Eqn. (\ref{est}) with an exponential moving average for some $\eta\in(0,1)$ forgetting rate:
\begin{equation}\label{adapt}
  a_j(t+1)=\eta\, a_j(t) + (1-\eta)f_j(y_t)=a_j(t) + (1-\eta) (f_j(y_t)-a_j(t)).
\end{equation}
We estimate density $\rho_t(y)= \sum_{j\in B} a_j(t)\ f_j(y)$ for a
given time $t$ based only on previous values, with exponentially
weakening weights $\propto\eta^{\Delta t}=e^{\ln(\eta)\,\Delta t}$ for
value $\Delta t$ time ago, allowing us to interpret $-1/\ln(\eta)$ as
characteristic lifetime.

There remains a difficult question of choosing the rate $\eta$, which
defines the strength of updates or lifetime, and which generally could
also evolve. To find the optimal value of $\eta$ (fixed here), we
searched the space of $\eta=0,0.01,\ldots,1$ for a fixed basis
$B=\{0,1,2,3,4\}$, evaluating log-likelihoods: the average
$\ln(\rho_t(y_t))$ for $\rho_t(y)=\sum_j a_j(t)\, f_j(y)$. However,
the problem is that values of $\rho$, as a polynomial, sometimes get below
zero; hence we need to reinterpret such negative predicted densities
as small positives, in what is referred to as calibration:  we instead used the
log-likelihood as an average $\ln(\tilde{\rho}_t(y_t))$, where
$\tilde{\rho}_t=\max(\rho_t,\epsilon)/N$ and $N$ is a normalization
constant to integrate to 1, and $\epsilon$ was arbitrarily chosen as
0.3 here.

\begin{figure*}[t!]
\begin{center}
\includegraphics[width=0.95\textwidth]{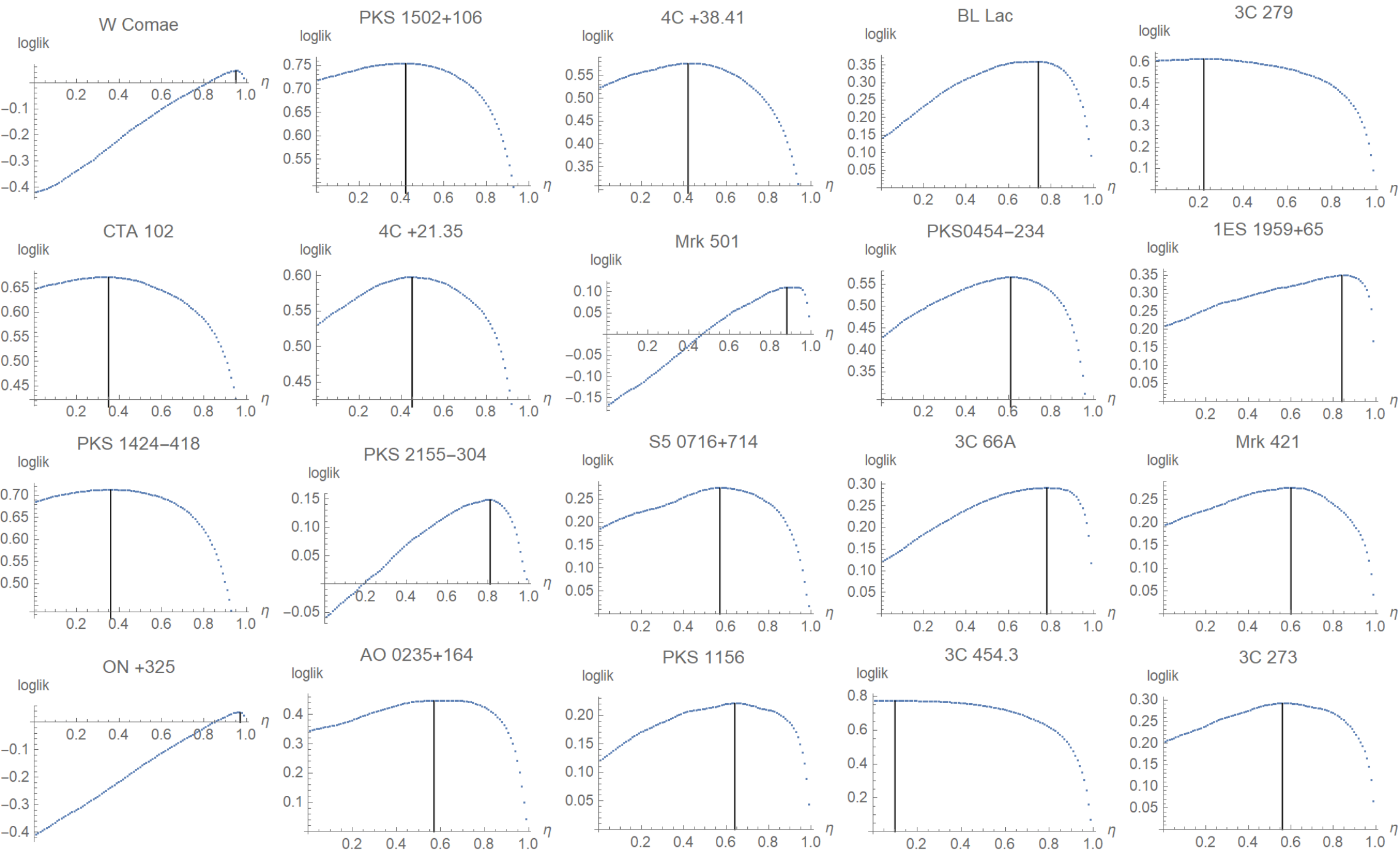}
\caption{Search for optimal values of the $\eta$ parameter (independently
    for each source) for evaluation of non-stationarity: while the 
  log-stable distribution obtained is averaged over a long period, the local
  probability distribution might evolve in time. For each normalized
  variable $y$ we found the time evolution of the coefficients $a_1, a_2, a_3, {\rm and} a_4$
  using an exponential moving average: $a_{j}(t+1)=\eta\,
  a_{j}(t)+(1-\eta) f_j(y_t)$.  The plots show the dependence of log-likelihood
  on $\eta$, without with the log-likelihood would be
  zero. We can see that in all but two cases (W~Comae, and ON~+325),
  we can essentially increase the log-likelihood by adjusting the
  parameters. Moreover, while e.g., in financial time series 
  $\eta$ is usually $>0.99$, here the optimal values of $\eta$ can be much smaller,
  corresponding to extremely fast forgetting of these systems. Both
  the optimal value of $\eta$ (e.g. interpreted as lifetime following $-1/\ln(\eta)$) and
  the corresponding log-likelihood can be used as features of the object
  describing non-stationarity for example for classification
  purposes.}
\label{fig:3}
\end{center}
\end{figure*}

The results of such a search for optimal values of $\eta$ using log-likelihood
evaluations are presented in Fig. \ref{fig:3}. While in financial time
series, optimal values of $\eta$ are usually close to 1, here they can be very far from 1,
suggesting strong non-stationarity --- that is, this optimal value of $\eta$ can be
treated as a feature characterizing non-stationarity of an object. The
log-likelihoods obtained are relatively large: while a stationary density
$\rho=1$ would have log-likelihood equal to 0, here the values can go up to
$\approx 0.8$, corresponding to mean values $\exp(0.8)\approx 2.2$ times 
localization in the range $[0,1]$.

\begin{figure*}
\begin{center}
\includegraphics{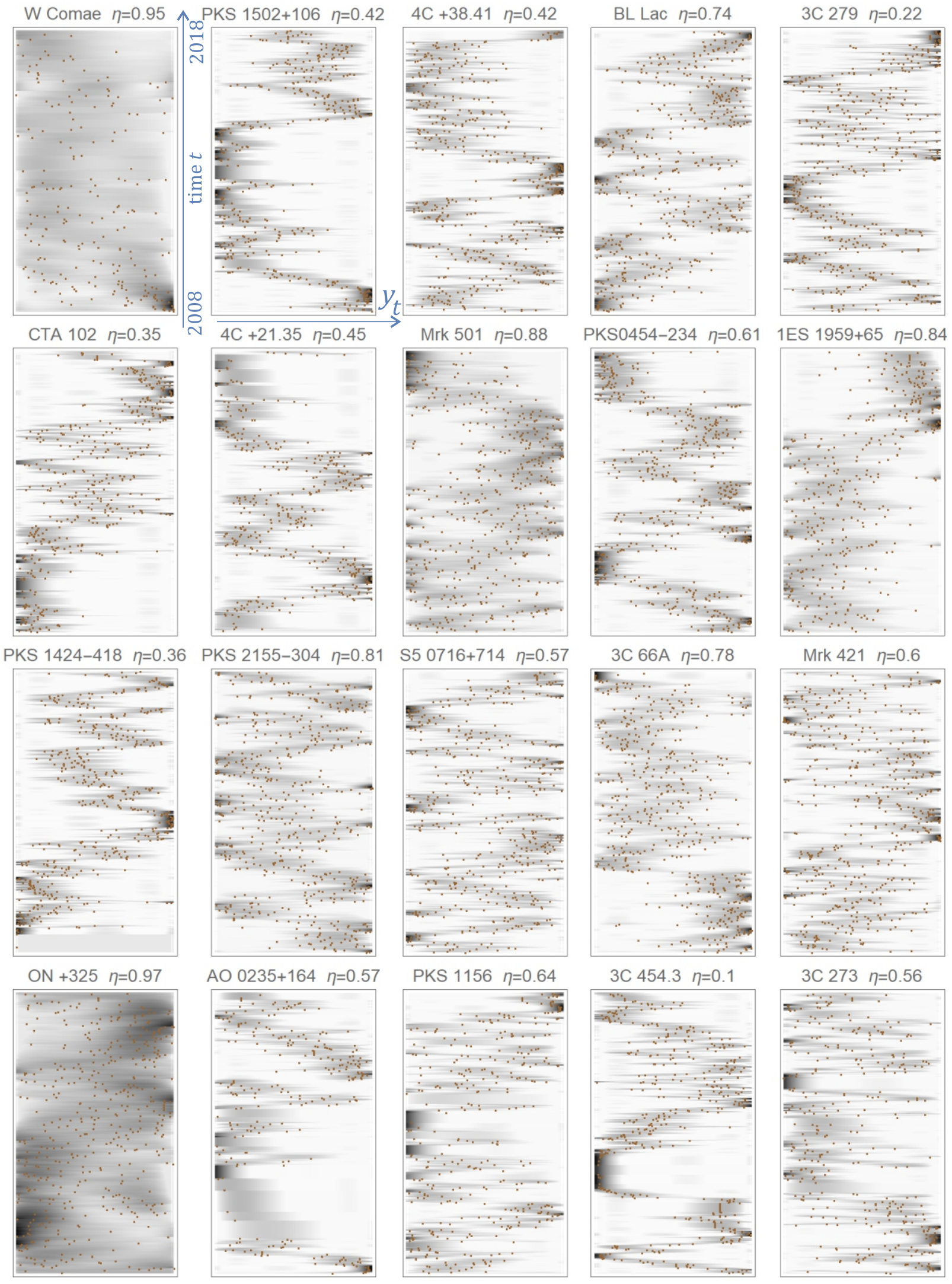}
\caption{Time-evolution of the estimated normalized
    density. In each plot the maximum likelihood forgetting rate
    obtained in Fig.~\ref{fig:3} was used.  The x-axis shows the normalized
    density range $[0,1]$.  The y-axis is time from bottom to top
    (2008--2018, synchronized for all objects). The points indicate the
    estimated normalized flux, and if projected to the x-axis would
    correspond to a nearly uniform density. The grayscale values represent
    the model density given by, $\rho(y,t) = 1 + \sum_{j=1}^4 f_j (y) a_j(t)$ for the values of $\eta$ optimizing the MLE
    (maxima in Fig.~\ref{fig:3}). The density  uses the evolution $a_j(t+1)=\eta
    a_j(t)+(1-\eta)f_j(y_t)$,  and its moments are denoted by  $j=\{1,2,3,4\}$. The gaps due to some of the missing values are filled with a constant density. Note that for some objects, such as 3C~279, PKS~2155$-$304, 
    and Mrk 421, we can directly see oscillations.}
\label{fig:4}
\end{center}
\end{figure*}

\begin{figure*}
\begin{center}
\includegraphics[width=0.9\textwidth]{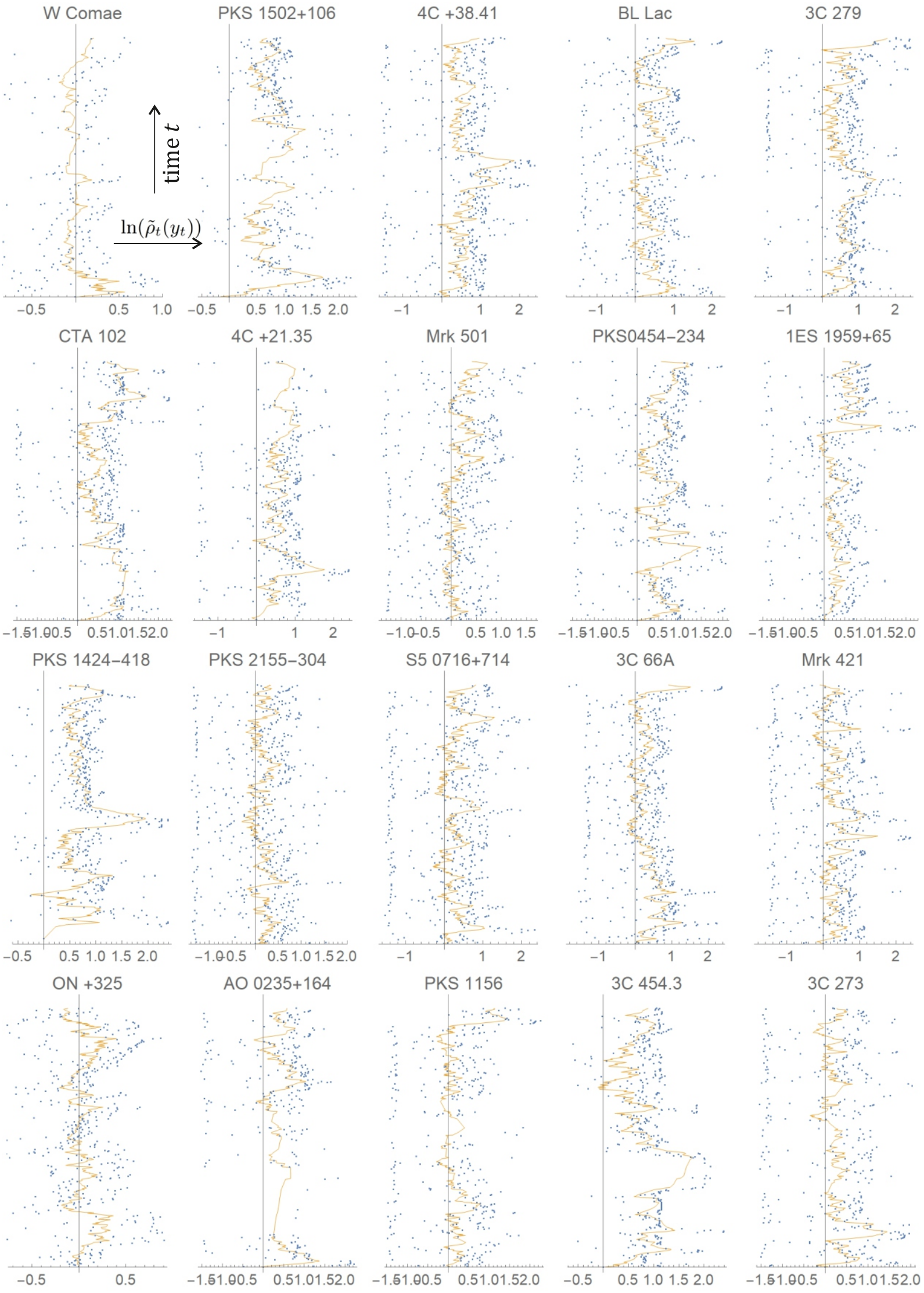}
\caption{The blue points show $\ln(\tilde{\rho}_t(y_t))$ in the horizontal
  direction and time evolution in the vertical direction.
  They average to log-likelihoods (maxima in Fig.~\ref{fig:3}),
  additionally allowing for evaluation of local time variability. The orange
  lines represent smoothing with an exponential moving
  average with rate 0.9. They can interpreted as local agreement with the models found;
  rapid decreases can be interpreted as anomalies.}
\label{fig:4.5}
\end{center}
\end{figure*}

Finally the predicted evolving densities using the optimized values of
$\eta$ are presented in Fig.~\ref{fig:4}, together with the $(y_t)$
points. Their values (horizontal direction) average to a nearly uniform
distribution; however, there are obvious clusters in their time
evolution (vertical direction), exploited in the adaptive model discussed here, and 
with predictions visualized as density. Figs.~\ref{fig:4.5} show the time
non-uniformity of evaluation of such predictions: $\tilde{\rho}_t(y_t)$
sequences (blue points), which smoothing (orange line), can be used to
evaluate local variability.  

The approach discussed here is optimized for fixed time differences
between measured values, which is not exactly true forthe data
analyzed here: we can see in Fig.~\ref{fig:4} that density is constant
between succeeding observations. Varying time difference could be
included, e.g., by faster modification (lower $\eta$) for longer time
differences, but such attempts did not lead to significant improvement
of evaluation, and hence are not presented.

We could also use separate variables $\eta_j$ for each parameter $a_j$ and
optimize them individually (also varying in time), modify the basis size and
$\epsilon$ in calibration; we performed such initial tests,
but we obtained nearly negligible improvement, so such tests are omitted
here for simplicity. The results presented also did not include errors
of values. They could be included, e.g., by replacing values with
discretized sets of values weighted with probabilities ($f_j(y)\to\sum
\textrm{Pr}(y) f_j(y)$), but such changes also yielded nearly
negligible impact on results.

\subsection{Autocorrelation analysis}

We also performed autocorrelation analysis for each series $(y_t)_{t=1..n}$
using a polynomial $(f_j)$ basis similar to \cite{Duda2021}. We
looked at pairs $(y_t,y_{t+l})$ shifted by lag $l$, up to a maximal lag
$m$ which was chosen here to be $m=100$ (weeks):
\begin{equation}
  P_l=\{(y_t,y_{t+l}): \textrm{ values in }t, t+l\textrm{ are available}\} \qquad \textrm{for}\qquad l=1,\ldots,m
\end{equation}
The data available have a regular time difference (7 days);
however, some values are missing. Hence we used all available
pairs with a chosen lag, and such sets of pairs usually have size varying
with lag (usually decreasing).

\begin{figure*}
\begin{center}
\includegraphics{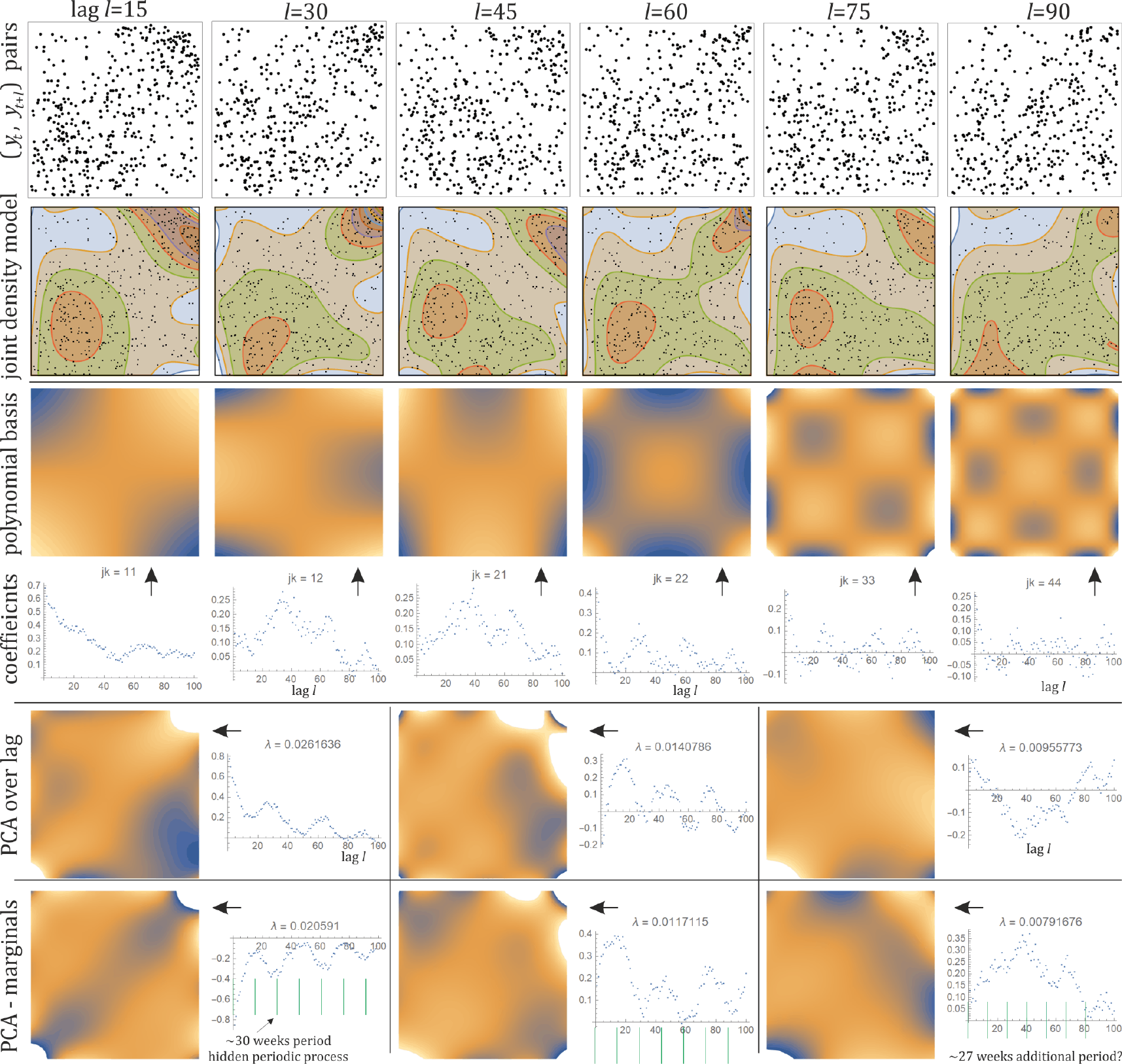}

\caption{Example of the autocorrelation analysis discussed for 3C~66A
  after normalization to nearly uniform variables
  $y_t=\textrm{CDF}(\ln(x_t))$. Row 1: $(y_t, y_{t+l})$ pairs for
  various lags $l$. Row 2: their joint distribution modeled as
  $\rho(y,z)=1+\sum_{jk} a_{j k}\, f_{j}(y) f_k(z)$. Row 3: densities
  (orange are positive, blue are negative) of some the 
  functions from the basis of orthonormal polynomials ($f_{j}(y)f_{k}(z)$)
  used for $(j,k)=(1,1), (1,2), (2,1), (2,2), (3,3), {\rm and} (4,4)$. 
  Row 4: their corresponding
  coefficients $a_{j k}(l)$ for various lags $l=1,\ldots,100$. Row 5:
  bases found with principal component analysis (PCA),
  the densities of the corresponding functions, and their
  eigenvalues.  Row 6: same as in row 5, but with earlier removal of
  contribution of marginal distributions
  $\tilde{a}_{jk}=a_{jk}-a_{j0}\,a_{0k}$, getting clearer signals only
  from dependencies between values shifted by lag $l$.  This way we get
  decorrelated multiple lag dependencies for each object; we
  present a manual attempt at fitting alternating minima-maxima that
  suggest periodic processes.}
\label{fig:5}
\end{center}
\end{figure*}

If uncorrelated, thanks to normalization, these pairs would be from
a nearly uniform $\rho=1$ joint distribution on $[0,1]^2$. We would like
to model distortion from this uniform distribution using a polynomial
basis. Let us start with the product basis of orthornormal polynomials
$(f_j(y)\cdot f_k(z))_{(j,k)\in B}$:
\begin{equation}
  \rho_l(y,z)=1+\sum_{(j,k)\in B^+} a_{jk}(l)\ f_j(y)\, f_k(z)
\end{equation}
Thanks to orthonormality we can use MSE estimation as in (\ref{est}):
\begin{equation}
  a_{jk}(l)=\frac{1}{|P_l|}\sum_{(y,z)\in P_l} f_j(y)\,f_k(z)
\end{equation}
As with $f_0=1$, the coefficients $a_{j0}$ describe marginal distributions of
the first variable, averaged over the second variable. Coefficients
$a_{0k}$ describe the marginal distribution of the second variable. $a_{11}$ is
the approximate dependence between their expected values, and has similar
interpretation as the correlation coefficient. Furthermore, the 
coefficients $a_{kl}$ can be viewed as higher mixed moments; they describe the dependence
between the $j$-th moment of the first variable and the $k$-th moment of the
second variable. Their direct interpretation is through the
density $f_j(y)f_k(z)$, which is presented in the third row of Fig.~\ref{fig:5}.

The top row of Fig. \ref{fig:5} contains examples of such pairs for
3C~66A. In the second row, we add isolines of joint density modeled
using the $B=\{(j,k):j,k=0,\ldots,4\}$ polynomial basis.  The third
row presents densities $f_j(y)f_k(z)$ for some $(j,k)$ corresponding
to $(1,1), (1,2), (2,1), (2,2), (3,3), {\rm and} (4,4)$. The fourth row shows the lag
$l$ dependence $a_{jk}(l)$ for these six presented coefficients.

In the last two rows, we try to improve upon the above arbitrarily
chosen basis by extracting features using PCA (principal component
analysis) over lag $l$. Specifically, for each object we have $|B|=25$
sequences for $m=100$ lags. Averaging over lags we can find the
$25\times 25$ covariance matrix $C_{j,k}$ and we can look at its few
eigenvectors corresponding to the highest eigenvalues: $C v = \lambda
v$. Then we define a new basis $f_v = v\cdot (f_{jk}:(j,k)\in B)$ and
corresponding sequence $a_v(l) = v\cdot (a_{jk}(l):(j,k)\in B)$ over
lag $l$. For the three highest eigenvalues we present the
contributions of $f_v$ to the joint density of $(y_t,y_{t+l})$.  

However, as discussed, these are highly non-stationary time series. If
we wanted to focus here on statistical dependencies of values shifted
by $l$, it would be beneficial to try to remove contributions from
non-stationarity. There are many ways to realize this; for example we
could use the evolved density modeled from non-stationarity analysis
for the additional normalization, but such analysis would be
model-dependent.

We used a simpler, more unequivocal approach instead: we subtracted 
the contributions of the (evolving) marginal distributions from the mixing terms:
\begin{equation}
  \tilde{a}_{jk}=a_{jk}-a_{j0}\,a_{0k}\qquad \textrm{for}\quad k,j>0.
\end{equation}

\begin{figure*}
\begin{center}
\includegraphics[width=0.91\textwidth]{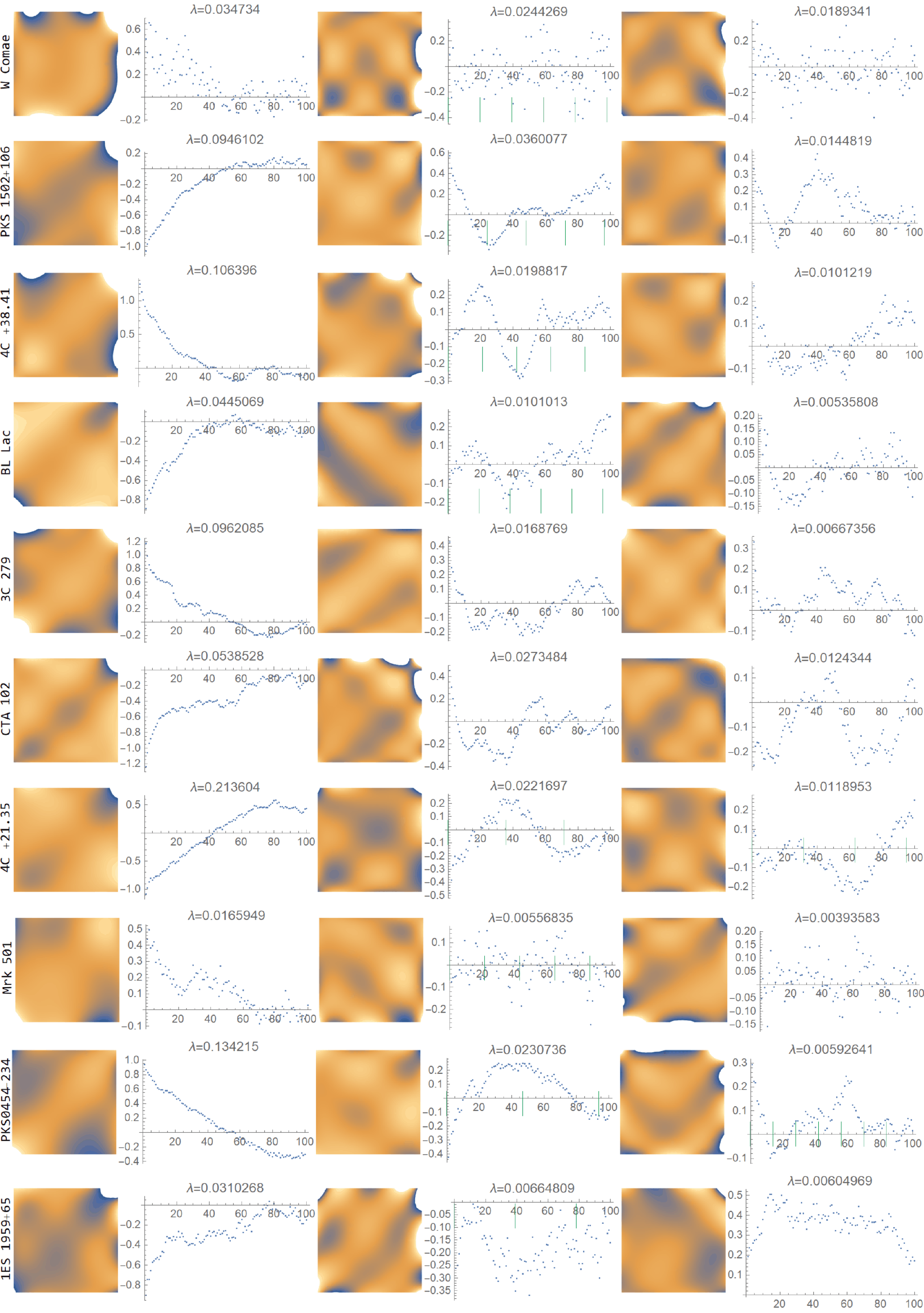}
\caption{Final plots for the first 10 sequences: PCA with removed
  marginal constributions as in row~6 of Fig.~\ref{fig:5}. Clear minima
  and maxima in lag dependence can be interpreted as characteristic
  timescales for a given object, with statistical interpretations presented in
  corresponding perturbations to joint densities
  $\rho(y_t,y_{t+l})\approx 1 +\sum_v a_v(l) f_v(y_t,y_{t+l})$. Green
  lattices present attempts to manually deduce periodic processes as
  alternating minima/maxima in fixed distances, to be improved in
  future work. }
\label{fig:6}

\end{center}
\end{figure*}

\begin{figure*}
\begin{center}
\includegraphics[width=0.93\textwidth]{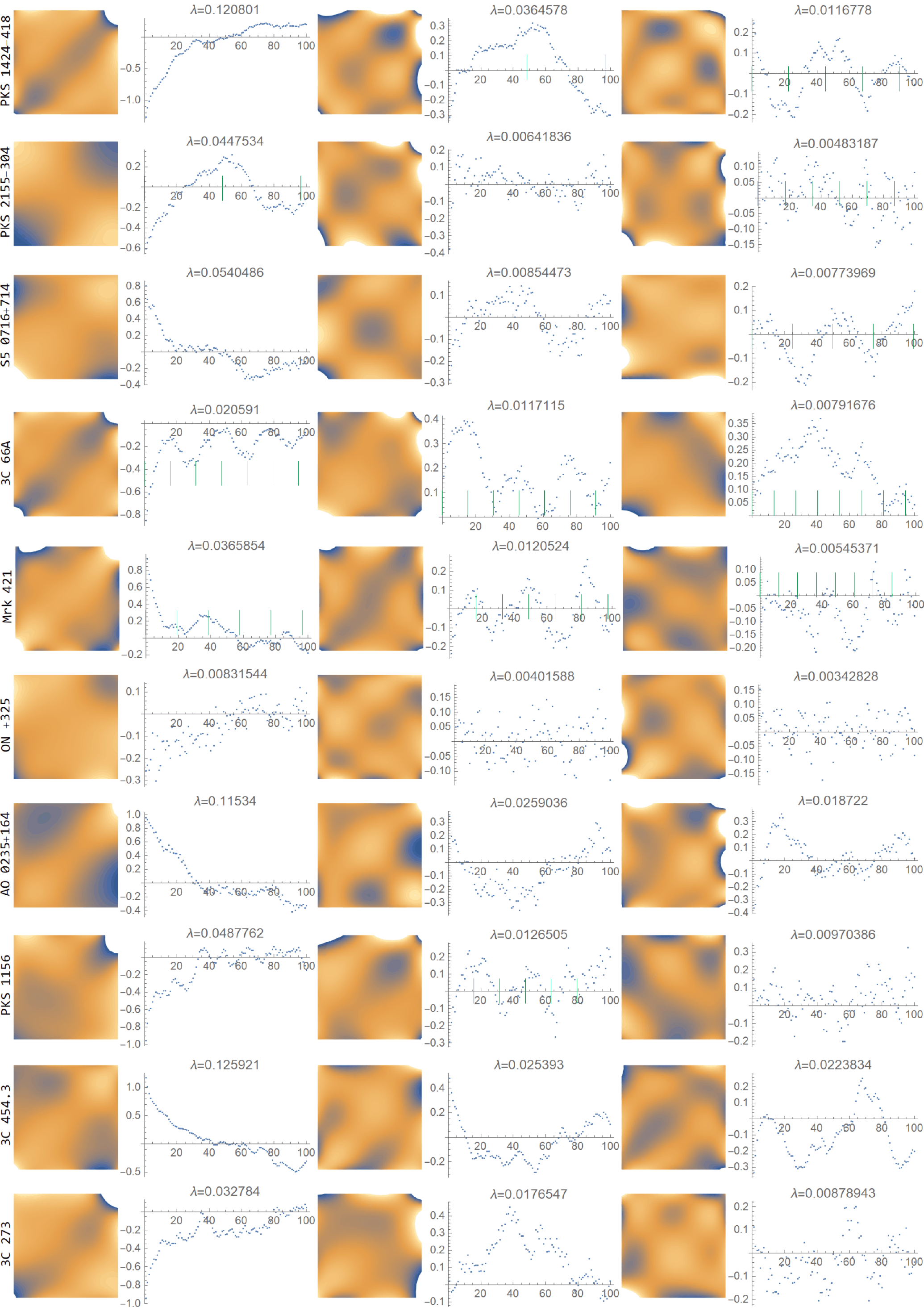}
\caption{Final plots for the last 10 sequences: PCA with removed marginal constributions 
as in row~6 of Fig.~\ref{fig:5}.}
\label{fig:7}
\end{center}
\end{figure*}

Then we could analogously perform PCA on $\tilde{a}_{jk}(l)$, leading
to results presented in the bottom row of Fig.~\ref{fig:5} and then analogously for all 20
objects in Figs.~\ref{fig:6} and \ref{fig:7}.

This way we obtain a few lag dependencies for each object, likely
nearly independent thanks to PCA. As we can see in Fig.~\ref{fig:6},
\ref{fig:7} they show complex features, but often have clear
minima-maxima structures, which could correspond to some characteristic
time differences. Their deeper analysis might be an involved work and
is planned for future research project; for example one could try to fit it with such
a dependence for coupled pendulums. A first suggestion is that periodic
processes should have alternating maxima/minima in fixed
distances. Green lines show the results of such manual attempts, to be
improved in a future work.

\section{Results and Discussion }
\label{sec:3}

We carried out time series analysis on the \gama-ray light curves of a
sample of 20 blazars, employing several methods that aimed to
constrain the statistical variability properties underlying the
decade-long observations. In particular, the variable blazar flux
distribution was investigated using log-stable PDFs. In addition, the
light curves were analyzed using novel methods dealing with the
statistical properties such as non-stationarity and autocorrelation.
In this section, we discuss the results of the analyses and attempt to
interpret them within the context of standard models of blazar
variability and emission.

\begin{itemize}

\item The blazar \gama-ray flux time series were fitted with general
  log-stable distributions parameterized by four parameters: location,
  variance, stability and asymmetry.  The maximum
  likelihood estimation for most of the sources result in $\alpha$=2,
  which indicates a log-normal distribution consistent with our previous
  result in \citet[][also see the references therein]{Bhatta2020}.
  Indeed a log-normal-like heavy-tailed flux distribution together with
  the linear RMS-flux relation could be a strong indication of the fact that the
  observed variability is driven by the multiplicative processes, which
  are non-linearly coupled over the large-scale jets. This then
  places constraints on the underlying mechanism producing the
  observed variable features in the light curves. For example, it 
  likely rules out the role of independent shot-noise-like processes
  as they represent additive processes.  In such a scenario, via a
  strong disk-jet connection, it is possible for the disk-based
  variations to make their way into the jet, propagate along the jet, and
  finally detected by the observer
  \citep[e. g. see][]{Giebels2009}. In the mini-jets scenario
  \citep[see][]{Biteau 2012}, a power-law flux distribution can result
  owing to the transformation of the isotropic distribution of the
  boosts (of the mini-jets) in the frame of reference as they make a small
  angle to the line of sight.

The analysis also revealed $\alpha<2$ heavy tails for some sources
($\rho(x)\sim |x|^{-\alpha-1}$), also an indication of multiplicative
processes of infinite variance. This could be an important result that
provides insights into the nature of \gama-ray production in blazars,
with an implication that the flux contributing to the higher end of
the heavier tail of the PDF probably consists of large amplitude
flaring events that could be of an different origin compared to the
origin of the lower amplitude fluxes. Moreover, in practice it is not
possible for the physical processes to realize infinite variances;
this suggests that there could be a cut-off at the higher end of the
flux distribution. Searching for the signatures of such cut-offs would be
important because it places a strong constraint on the characteristic
highest energies attainable in these systems. Such a characteristic
energy would in turn shed light on the nature of the dominant particle
acceleration scenario, e.g., shock waves and magnetic reconnection,
contributing to the \gama-ray production in the jets.

\item One of the important results that the work revealed is that,
  although the PDF follows a log-stable distribution over a long period, it
  can display transient non-stationarity features, suggesting that the
  processes linked to the origin of variability are fast-forgetting. 
  Such non-stationarity features, in the form of the fast
  changing PDFs, can be considered as statistical fluctuations in
  the long-term trend owing to local MHD instabilities either in the
  disc or in the turbulent jets
  \citep[e.g.][]{Calafut2015,Marscher14}. Recently, transient
  non-stationarity in the form of variable PSD slopes was also
  reported in the X-ray observations of blazars, which were found to be
  consistent with short memory processes
  \citep{Zhang2021,Bhattacharyya2020}. The main parameters of the
  observed non-stationary are the forgetting rate $\eta$, which can be linked
  to lifetime $-1/\ln(\eta)$, and log-likelihood, which describes
  strength of localization.  These parameters can represent some of
  the characteristics of intrinsic dynamics of the instability
  events. In addition, such quantities can be incorporated in the scheme
  of source classification.

\item Novel auto-correlation analysis exploring the lag dependence of
  multiple mixed moments shows complex minima/maxima structures, and
  the timescales corresponding to these extrema are typically on the
  order of a few months.  Recurrence analysis performed on the
  observations resulted in similar characteristic timescales, 
  so-called trapping timescales, as reported in \citet{Bhatta2020a}. The
  timescales can be interpreted as some characteristic timescales
  associated with the jet processes. These timescales could be driven
  by the accretion disk-related timescales e.g., dynamical, thermal, or
  viscous timescales \citep[see][]{Czerny2006} in an AGN with a
  central black hole of mass on the order of$\sim 10^{8}-10^{9}
  M_{\odot}$; however, the timescales could be altered by the jet
  Lorentz factors. Additionally, the observed timescales can also be
  linked to Rayleigh-Taylor and Kelvin-Helmholtz instabilities
  developing at a disk-magnetosphere interface \citep{Li2004}, or
  non-thermal (e.g. synchrotron and inverse-Compton) cooling
  timescales of the accelerated charged particle in the jet. Apart
  from aperiodic timescales, various jet and accretion disk-related
  instabilities can set up (quasi-) periodic oscillations as observed
  in the multi-frequency light curves of several blazars
  \citep[see][and references therein]{Bhatta2019}.
  
  \item Finally, it should be stressed that, although some of the results obtained here e. g., transient non-stationary features and heavy-tailed distribution, may have found a convenient interpretations  in terms multiplicative and non-linear processes occurring at the turbulent relativistic jets,  it is possible a broad range of stochastic processes to produce statistical properties similar to the observed time series, e. g., RMS-flux relation and log-normality. This makes it difficult to arrive at a common understanding of the physical model that drives the observed blazar variability. Furthermore, it appears that a robust correspondence between the observed properties, and the linearity/non-linearity and additive/multiplicative is yet to established \citep[see][for a detailed discussion on the topic]{Scargle2020}.    

\end{itemize}

\section*{Acknowledgments}
The authors would like to thank the anonymous referee for his/her
useful comments and suggestions which help improve the quality of the
paper. GB acknowledges the financial support by Narodowe Centrum Nauki
(NCN) grant UMO-2017/26/D/ST9/01178.  The authors thank Alex Markowitz
for improving the English grammar and clarity of this manuscript.

\section*{Data availability}
The raw \textit{Fermi}-LAT data used in this article can be accessed
from the LAT server
(\url{https://fermi.gsfc.nasa.gov/cgi-bin/ssc/LAT/LATDataQuery.cgi}). Furthermore,
the processed data will be shared on reasonable request to the
corresponding author.

\end{document}